# Scrutinizing uncitedness of selective Indian physics and astronomy journals through the prism of some h-type indicators

Amit Kumar Das[a] and Bidyarthi Dutta[b]

[a]Central Library, Bhatter College, Dantan, Paschim Medinipur - 721 426, West Bengal; email: amitkumardas19@yahoo.co.in
[b]Department of Library and Information Science, Vidyasagar University, Midnapore 721 102, West Bengal; email: bidyarthi.bhaswati@gmail.com

## Abstract

The citation analysis is the central focus of metric studies, may it be bibliometrics, scientometrics or informetrics. It deals with objects received citations in due course of time. But there are huge chunk of academic items receiving no citation years after years and remaining beyond the veil of ignorance of the academic audience. These are known as uncited items. Now, the question is, why a paper fails to get citation? The attribute of incapability of receiving citation may be termed as *Uncitedness*. This paper traces brief history of the concept of uncitedness sprouted first in 1964 in an article entitled *Cybernetics, homeostasis and a model of disease* by Gerson Jacobs. The concept of uncitedness was scientometrically first explained by Garfield in 1970. The uncitedness of twelve esteemed Indian physics and astronomy journals over a twelve years' (2009-2020) time span is analysed here. Besides Uncitedness Factor (UF), three other indicators are introduced here, viz. Citation per paper per Year (CY), h-core Density (HD) and Time-normalised h-index (TH). The journal-wise variational patterns of these four indicators, i.e. UF, CY, HD and TH and the relationships of UF with other three indicators are analysed. The calculated numerical values of these indicators are observed to formulate seven hypotheses, which are tested by F-Test method. The average annual rate of change of uncited paper is found 67% of total number of papers. The indicator CY is found temporally constant. The indicator HD is found nearly constant journal-wise over the entire time span, while the indicator TH is found nearly constant for all journals. The UF inversely varies with CY and TH for the journals and directly varies with TH over the years. Except few highly reputed Indian journals in physics and astronomy, majority other journals face the situation of uncitedness. The uncitedness of Indian journals in this field outshines the same for global journals by 12%, which indicates lack of circulation and timely reach of research communication to the relevant audience.

## Introduction

The studies on scientometrics or bibliometrics always focus on highly cited items, while a hefty chunk of publication output remains either uncited or very feebly, only once, twice, or thrice cited, over the years. The literal meaning of the word 'uncited' is 'not quoted' or 'not cited', which is just opposite to 'cited'. The story of uncitedness has its roots that trace back to more than half a century ago. Perhaps the article entitled *Cybernetics, homeostasis and a model of disease* by Gerson Jacobs[1] (1964) was the starting signal behind the blooming of the concept of uncitedness. Despite a large number of reprint requests shortly after the publication of this article, the same was indexed neither in any bibliography nor in *Science Citation Index*, as complained by the Author seven years later in the *Journal of American Medical Association*[2] under *Letters to the Editor* section. The Author, Gerson Jacobs further pointed out five probable reasons for the same[2], though he identified the first one as the dominating factor, which states that *the article is too profound and difficult to understand for the present generation of scientists*. The second and third reasons for the non-inclusion of the article, though are also very interesting. The second reason stated that *the article is a threat to the establishment*, while the third reason stated *it is not radical chic to cite an article that has never been cited*. The second reason however says a pertinent cause of negative citation. The third reason unveils the actual cause behind the very centripetal nature of citation accumulation, which is the basis for the well-known Preferential Attachment Model or Mathew Effect. This reason emphasizes the fact that citation has always a tendency to follow some precursors resulting in the common feature of citation attracts citation. This centripetal nature of citation, in turn, explains why a bulky chunk of articles remains uncited over the years. Jacob, the author put forth an exciting suggestion here, that is *the Science Citation Index should establish a section "The never cited index". This will allow the truly relevant scientist to search out the literature of the revolutionary and suppressed literature*. The phrase *radical chic* in the third reason came from the famous 1970's book by Tom Wolfe entitled *Radical Chic & Mau-Mauing the Flak Catchers*[3]. This phrase has entered into the socio-political and socio-cultural glossary to describe the approval of radical or quasi-radical causes by members of the elite class society. It seems G. Jacob, the author used this phrase bit sarcastically in the third reason indicating elitism behind the receiving of citation.

Jacob's letter was acutely criticized by Garfield[4], where he directly alleged that Jacob is mistaken on two points, at first his article was abstracted in *The Journal* and its citations appeared in the *Science*

*Citation Index* of 1964. Secondly, his idea of a Never-Cited Index or *Index Oblivionis* was not an original concept. Garfield pointed out that information on uncitedness first appeared in *Genetics Citation Index*[5] in 1963. The first explicit use of the word 'Uncited' was found by Garfield[6] in 1970. In this paper, Garfield opined that many uncited papers may be an excellent source of material for graduate students. In another contemporary paper[7], Garfield opined that obsolescence was the relatable reason for the continuous growth of uncitedness of research articles. He said that *I am constantly frustrated by the fact that citation indexes in most fields are not yet available for the first sixty years of the Twentieth Century.* This saying pondered the discipline-wise non-uniformity amongst the citation-accumulation patterns. Garfield, in this paper, reminded us, it is the duty of the librarians to assist the stakeholders in the selection of their thesis and dissertation topics by identifying interesting but hitherto uncited articles and bringing them to the focal point subsequently.

## Literature Review

Ghosh[8,9] studied the uncitedness of 222 articles published in the *Journal of the American Chemical Society* from January to February 1965 and concluded that on average, 14.7% of articles remained uncited during any given year. Lawani[10] showed that the rate of uncitedness declined with the increasing quality of articles for cancer literature. Stern[11] identified some bibliographic characteristics that distinguished cited papers from uncited papers. Sengupta and Henzler[12] analyzed time lag between publications, average citation time, and uncitedness of cancer literature. Szava-Kovats[13] analyzed the nature and phenomenology of non-SCI eponymous citedness of physics literature. Hamilton[14] reported that on average, 47.4%, 74.7%, and 98% of articles remain uncited over five years in the disciplines of science, social science, and arts & humanities respectively. He also pointed out the wide variations of uncitedness for different subjects even within a discipline. For instance, only 9.2% of articles remained uncited in the field of atomic, molecular, and chemical physics followed by virology (14.0%), physics (16.7%), organic chemistry (18.6%), etc. While in acoustics, the percentage of uncitedness was 40.1% followed by optics (49.1%), developmental biology (61.5%), and electrochemistry (64.6%). In engineering science, every subject field showed high rates of uncitedness, with civil engineering highest at 78.0% and biomedical engineering (59.1%) figured lowest. Hamilton[15] pointed out that these figures of uncitedness were obtained from the statistics of Garfield's *Institute for Scientific Information (ISI)*, while *ISI*'s database then covered some 4500 (only 6%) out of nearly' 74,000 scientific journal titles listed in Bowker/ Ulrich's database. However, Pendlebury[16] explained the high percentage of uncitedness of SCI-journals. The SCI journals contained not only articles but also other forms of documents like reviews, notes, meeting abstracts, editorials, obituaries, letters, etc. which were, by and large, remained uncited. Pendelbury[16] defined uncitedness from the viewpoint of ISI's journal coverage. Garfield[17], however, differed from Hamilton's explanation of uncitedness and opined that *due to the cumulative character of science and scholarship, a great deal of the literature is cited but once*. Garfield coined the term *Onesies* to indicate the once-cited papers and found out in a study for the years 1945-88 that nearly 56% of publications remained *Onesies*.

Schwartz[18] found out the large-scale uncitedness percentage for library and information science, which figured 72%. Van Dalen and Kene[19] found that after ten years 24% of the demography articles were still uncited with an average number of citations per article figured seven. Small[20] introduced a normative theory of citation viewing the same as symbolic payment of intellectual debts. He coined the term *citationology* as a subject domain to embrace all aspects of studies related to citedness and uncitedness within its periphery. Leeuwen, Thed and Moed[21] et al showed the inter-relations among the journal impact factor, degree of uncitedness, citation frequency distribution, and output of a volume. Van Dalen, Hendrik and Henkens[22] studied 1371 articles published in 17 demography journals during 1990-92 and concluded that the state of uncitedness did not affect the future probability of being cited. Egghe[23] showed the impact factor as a decreasing function of the uncitedness factor. Onyancha[24] compared the performance of 13 library and information science journals using their citedness and uncitedness along with other indicators like the number of citations, h-index and g-index, etc. Wallace, Vincent and Yves[25] proposed a simple model based on a random selection process to explain the "uncitedness" phenomenon and its decline over the years based on the Web of Science.

Egghe[26] found out the functional relation between the impact factor and the uncitedness factor based on Central Limit Theorem. Egghe, Guns and Rousseau[27] presented an interesting finding that the Nobel laureates and Fields medallists in the fields of physics, chemistry, physiology, or medicine and mathematics (field medallists) have a rather large fraction (10% or more) of uncited publications. The most remarkable result here was a positive correlation between the h-index and the number of uncited articles. Hsu and Ding-Wei[28] derived scaling relation between the impact factor and the uncited percentage by a random mechanism based on the cumulative advantage process. Burrell[29] argued that Egghe's[27] results might, at first sight, seem to be surprising but still explainable in a stochastic framework. Burrel[30] raised questions and discussed some of the arguments of Hsu's[28] and Egghe's[23,32] articles. Heneberg[31] analyzed the uncitedness among two independent groups of highly visible scientists, which included Fields medallist mathematicians and

Nobel laureate researchers in physiology or medicine. The result revealed that over 90% of the uncited database records of highly visible scientists can be explained by the inclusion of research output other than articles, i.e. editorial materials, meeting abstracts, letters to the editor, etc. and also by the errors of omission and commission of the Web of Science database and of the citing documents. Egghe[32] presented a heuristic proof of the relation between the impact factor (IF) and the uncitedness factor (U), the fraction of the uncited papers, i.e. $U = \frac{1}{1+IF}$. Law, Andy and Norman[33] analyzed the uncited articles published in the *Asia Pacific Journal of Tourism Research* and the *Journal of Travel & Tourism Marketing* during the period 1996-2005. Garg and Kumar[34] analyzed 35,640 papers published by Indian scientists in 2008, indexed by Science Citation Index-Expanded (SCI-E), which revealed that 6231 (17.5%) papers remained uncited during 2008-2013. The highest proportion of uncited papers was in the discipline of agricultural sciences followed by multidisciplinary and mathematical sciences. Lou and He[35] collected uncited papers from 24 journals in six subjects from WoS and found that there is a significant correlation between affiliation reputation and uncitedness. Arsenault and Vincent[36] found out a correlation between the uncitedness factor and alphabets in the authors' names.

Liang, Zhong and Rousseau[37] studied three types of uncitedness in Library and Information Science journals, viz. uncitedness for articles, authors, and topics. Gopalakrishnan, Bathrinarayanan and Tamizhchelvan[38] carried out a bibliometric study of uncited publications in "micro-electromechanical systems" literature. Elango[39] discussed the characteristics of uncitedness of literature on tribology and compared it with cited papers. The results showed there was a significant difference in characteristics between cited and uncited papers. Zewen and Yishan[40] identified seven major points, i.e. research hotspots and novel topics, research topics similar to one's work, high quality of content, reasonable self-citation, highlighted title, prestigious authors, and academic tastes and interests similar to one's own, that usually facilitate the easy citation of papers. Zewen, Yishan and Jianjun[41] considered the mutual relations and closeness degree between the non-citation factors and different influencing factors and found out that three variables, i.e. average number of authors per paper in the journal, the average number of references per paper in the journal, and issues of the journal did not exert an influence on the decline of percentages of never-cited papers in the citation time window. Zewen, Yishan and Jianjun[42] used a survey-based structural equation model and established that three observed variables of 'academic status of the journal' including 'public praise of journal', 'impact factor of the journal', and 'member of SCI, EI, and Scopus Journals', showed the highest values of indirect effect on the non-citation rate. Yeung[43] found relationships among various citation metrics in the field of neuroimaging. Nowroozzadeh and Salehi-Marzijarani[44] analyzed uncitedness in the top-ranked medical journals. Nicolaisen and Tove[45] showed large variation in uncitedness ratios between subject areas and also between document types in seven subject-area and seven document types. Baruch, Fabian and Abdul-Rahman[46] found from the analysis of a sample of 2777 papers in management studies that the rate of uncitedness is quite low, only 6.5% in this field. Dorta-González, Rafael and María[47] analyzed three factors for journals, conference proceedings and book series, i.e. the subject field, the access modality (open access vs. paywalled) and the visibility of the source. They found no strong correlation between open access and uncitedness, but lower uncited rates of open access journals.

## Research Gap

Hamilton[14] noticed an average of 47.4%, 74.7%, and 98% uncitedness in the disciplines of science, social science and arts & humanities respectively in 1991, while the picture has still remained unchanged even after 30 years. Lloyd and Ordorika[48] pointed out in 2021 that in Scopus, 49% of citations are of publications in the life sciences and medicine followed by the natural sciences (27%) and engineering and technology (17%). The social sciences and arts & humanities represent just 6% and 1% of citations. It is not only the discipline-wise large variation of uncitedness, but the uncitedness factor also shows the acute non-uniform pattern over journals, institutes, and even countries as reflected from the literature review. The notable point here is that the country-wise study of uncitedness is very few as found from the literature review. It is also noticed that only two articles[34,38] discussed the uncitedness of Indian scientists. The former[34] article highlighted the uncitedness of Indian scientists, while the later[38] one showed that 31.44% of Indian articles on micro-electro-mechanical systems (MEMS) remained uncited. Although, the uncitedness of Indian journals in almost all major disciplines is a key issue today, but no research in this domain has yet been observed that created a research gap. This paper analyses the Uncitedness Factor (UF) of major Indian physics and astronomy journals, which according to Egghe[23], is defined as the ratio of the number of uncited papers (U) to the total number of papers (P) of the respective journals in a particular year. The UF thus figures out the fractional change in uncited papers with respect to the total number of papers, i.e. $\mathrm{UF} = \frac{\mathrm{U}}{\mathrm{P}}$.....(1)

Besides Uncitedness Factor (UF), other three h-type indicators viz. CY, HD, and TH are defined here and their correlations with Uncitedness Factor (UF) are tested.

## h-type Indicators

### *Citation-per-Paper-per-Year (CY)*

Let 'P' number of articles published in a journal in the year 'Y' have received 'C' number of citations in the current year, $Y_c$ say. Then Citation-per-Paper-per-Year, denoted by 'CY' is defined as, $CY = \frac{C}{P(Y_c - Y)}$ .....(2)

### *h-Core Density (HD)*

Let 'P' number of articles published in the concerned journal in the year 'Y' have received 'C' number of citations in the current year, $Y_c$ say, where the value of h-index is h, say and $C_h$ denotes the number of h-core citations. The h-core Density, denoted by 'HD' is defined as, $HD = \frac{C_h}{C}$ .....(3)

### *Time-Normalized h Index (TH)*

Let 'P' number of articles published in the concerned journal in the year 'Y' with the value of h-index is 'h' in the current year $Y_c$ (say). The Time-Normalised h-Index, denoted by 'TH' is defined as, $TH = \frac{h}{(Y_c - Y)}$ .....(4)

## Research Question

1) How the uncitedness factor for Indian physics and astronomy journals varies with time?

2) How the uncitedness factor for Indian physics and astronomy journals varies with other h-type indicators, i.e. CY, HD and TH over a time window?

3) Is there any empirical relationship existing between UF and either of CY/ HD/ TH?

## Purpose of Study

1) To observe how do the citations influence the uncitedness of Indian physics and astronomy journals.

2) To observe how h-core citation and h-index influence uncitedness of Indian physics and astronomy journals.

3) To find out the numerical values of UF, CY, HD, and TH for 12 Indian physics and astronomy journals over 12 years of duration (2009-2020).

4) To represent the empirical relation between UF and either of CY, HD and TH in terms of the functional equation from the calculated data.

## Hypothesis Formulated

The following seven null hypotheses ($H_0$) have been formulated for this study. The first four hypotheses are about the constancy factor of the four indicators, while the last three hypotheses state the relationships of UF with other three indicators. The fundamental axioms forming the basis of these hypotheses are as follows:

**Axiom 1)** The preferential attachment or cumulative advantage model[49,50,51] of the citation accumulation process by any item (article, journal, author or institution) is the foremost axiom. This model explains wealth or credit distribution among a number of individuals or objects according to how much they already have, so that the wealthy ones or haves receive more than the have-nots. Similarly, the citation always has a tendency to accrue around the cited papers. The higher cited papers usually attract more citations, but majority uncited papers are seldom cited highly.

**Axiom 2)** The citation accumulation process is a function of time, i.e. citations gradually amass to an article as time passes on.

**Axiom 3)** The h-index is the solution of the equation: $r = C(r)$, where $C(r)$ is the number of the citations of the $r^{th}$ publication from the ranked list or articles of the researcher[52]. The h-core citation is thus a dependent function of total citation, as citation determines the rank.

**Hypothesis 1) $H_0(1)$:** The total number of papers is directly proportional to the number of uncited papers in a journal in any year, i.e.

$$P \propto U, \text{ or UF (Uncitedness Factor)} = \frac{U}{P}$$

$$= \text{Constant both over the years and over the journals as well . (Equation (1))}$$

**Hypothesis 2) $H_0(2)$:** The number of citations (C) is directly proportional to the total number of papers (P) in a journal/ by an author, and also directly proportional to the years spent after publication $(Y_C - Y)$, where $Y_C$ and $Y$ are the current year and the year of publication respectively.

$$\text{Hence, } C \propto P, \text{ when } (Y_C - Y) \text{ is constant}$$

$$C \propto (Y_C - Y) \text{ , when P constant}$$

$$C \propto P(Y_C - Y), \text{ when both vary}$$

The Citation-per-Paper-per-Year (CY) or $CY = \frac{C}{P(Y_C - Y)} = \text{Constant factor both over the years and over the journals as well.}$

**Hypothesis 3) $H_0(3)$:** The h-core citations ($C_h$) is directly proportional to the total number of citations (C) for a journal in any year, i.e. $C_h \propto C$, or $\frac{C_h}{C} = h - \text{core Density} = \text{constant over the years and over the journals as well.}$

**Hypothesis 4) $H_0(4)$:** As citation accumulation is a function of time or years spent, h-index increases with years passing on. Hence, $h \propto (Y_C - Y)$; or $\frac{h}{(Y_C - Y)} = $ Constant, or TH is constant over the years and over the journals as well.

**Hypothesis 5) $H_0(5)$:** UF is inversely proportional to CY, or $UF \propto \frac{1}{CY}$, or UF*CY = Constant over the years and over the journals as well.

**Hypothesis 6) $H_0(6.1)$:** UF is inversely proportional to TH, or $UF \propto \frac{1}{TH}$, or UF*TH = Constant over the journals and **$H_0(6.2)$:** TH is directly proportional to UF, or $TH \propto UF$, or TH/UF = Constant over the years.

**Hypothesis 7) $H_0(7)$:** HD is directly proportional to UF, or $HD \propto UF$, or HD/UF = Constant over the years and over the journals as well.

## Scope and Methodology

The values of the four indicators, viz. UF, CY, HD and TH are calculated for twelve esteemed Indian physics and astronomy journals from 2009 to 2020 (Appendix: Table A1 to Table A4) based on the available primary data (Appendix: Table A8). The average value in each year calculated journalwise and the average value of each journal calculated yearwise, represented by Mean(Y) and Mean(J) respectively, are furnished in the bottom-most row and extreme right column of Table A1 to Table A7. Of the twelve journals, seven journals belong to the core domain of physics and astronomy (S. No. 2, 4, 5, 6, 8, 11 and 12), while remaining five journals belong to allied interdisciplinary areas of physics but publish articles on physics regularly (S. No. 1, 3, 7, 9 and 10). The journal of Serial No. 10 belongs to the entire natural science discipline but publish physics articles on regular basis. It is very old and esteemed Indian science journal. The list of the said twelve esteemed Indian journals selected for this study is furnished below:

1) Defence Science Journal (DSJ)
2) Indian Journal of Biochemistry and Biophysics (IJBB)
3) Indian Journal of Engineering and Materials Sciences (IJEMS)
4) Indian Journal of Physics (IJP)
5) Indian Journal of Pure & Applied Physics (IJPAP)
6) Journal of Astrophysics and Astronomy (JAA)
7) Journal of Earth System Science (JESS)
8) Journal of Medical Physics (JMP)
9) Journal of Scientific and Industrial Research (JSIR)
10) Proceedings of the Indian National Science Academy (PINSA)
11) Pramana - Journal of Physics (PJP)

The primary data furnished in Table A8 have been collected from *Scopus* database. The search strategy followed in *Scopus* under 'Advanced Search' was, "SUBJAREA(PHYS) AND AFFILCOUNTRY (INDIA) AND (EXACTSRCTITLE(DEFENCE SCIENCE JOURNAL))". The time range was set from 2009 to 2020. The same strategy was repeated for the other eleven journals as listed above and the number of papers, number of uncited papers and total number of citations in each of the journals from 2009 to 2020 as retrieved from Scopus are presented in Table A8. The h-index and h-core citations for each of the journals are calculated from the retrieved data. The yearwise and journalwise breakup of the values of the indicators along with their relation with UF are presented in Table A1 to Table A7. The seven hypotheses formulated are tested by F-Test method and the results are presented in Table 1 (for journals) and Table 2 (for years). The six statistical parameters, viz. Mean, Median, Range, Standard Deviation, Coefficient of Variation and Kurtosis of the four indicators, along with the relations between UF and the remaining three indicators are presented in Table 3 (for journals) and Table 4 (for years) respectively. The correlation coefficients between UF-CY, UF-TH and UF-HD are presented in Table 3 (for journals) and Table 4 (for years) respectively.

## Results and Analysis

The primary data obtained from twelve journals listed above is furnished in Table A8 (Appendix). The total number of papers published in the respective journals (P) along with the number of uncited papers (U), total number of citation (TC) and h-index for twelve journals are listed here. In all, 13,567 papers are published in 12 journals from 2009 to 2020, which received 67,365 citations figuring 5 citations per paper on an average. Of the entire publications, 3,884 papers received no citations till date figuring 28.6% of uncited papers. The Change in total number of papers (P), time-normalised Total Citation (TC) and number of Uncited papers (U) from 2009 to 2020 is presented in Figure 1. The journal-wise variation of the total number of papers and the total number of uncited papers is presented in Figure 2. The variation of total number of citations over the journals is presented in Figure 3. The largest number of papers was published in IJP that figured 2368, followed by PJP (2356), JESS (1540) and IJPAP (1348). These four journals published 56% of entire publications. The highest citation was received by IJP (10714) followed by JESS (10702), PJP (10600) and IJPAP (7296). It is worthwhile to mention that IJP, started in 1925, is the oldest Indian physics journal right now. These four journals altogether received 39312 citations, i.e. almost 59% of total (67365) citations. The average citation per paper is highest for IJBB (8), followed by JESS (7) and JMP & JSIR (6 each). The highest number of uncited papers is found in PJP (709), followed by IJP (597) and PINSA (382). In terms of percentage of uncited papers, PINSA ranked first, where 50% papers remained uncited, followed by JAA (42%), PNASI (40%) IJEMS (31%). The journal

IJBB lowest uncited percentage (15%) followed by JESS (19%) and JMP (24%). The three highly reputed journals, viz. IJP, IJPAP and PJP found uncited percentages figured 25%, 28% and 30% respectively.

The yearwise result shows, the lowest and highest numbers of papers were found in 2009 (976) and 2020 (1508) with several intermittent fluctuations over the time span. Also, the lowest and highest numbers of normalised citations were found in 2019 (560) and 2020 (1058), the crest-trough pair just in consecutive years, which is very interesting figure. The lowest numbers of uncited papers were found in 2010 and 2013 that figured 141, i.e. 14% of total number of papers published in respective years. On the contrary, the largest number of uncited papers was found in 2020 (968), i.e. 64% of total papers of the year. The number of uncited papers gradually enhanced with years as citation accumulation is a function of time.

In all, The Uncitedness Factor (UF) does not show steadiness either for the journals or over the years as $H_0(1)$ is rejected both for the journals (Table 1) and for the years (Table 2). The journal PINSA possessed highest average UF (0.47) followed by JAA (0.38) and PNASI (0.37). Also, the highest value of the UF has been observed in the year 2020 (0.68) followed by 2019 (0.51), 2018 (0.37), 2017 (0.25), 2016 (0.24) and so on. The steady decreasing trend of the UF with years in reverse chronological order accords the negative power model, i.e. $UF = 0.67 * t^{-0.6}$, Coefficient of Determination $(R^2) = 0.931$ and 't' indicates time in years. The Coefficients of Variation (CV) of UF for journals and years figure 33.3% and 57.7% respectively, which are pretty high showing far from constancy tendency. Also the Kurtosis values for $(UF)_{Journal}$ and $(UF)_{Year}$ that figured 0.049 and 2.766 respectively show skewed patterns, particularly high positive value of $(UF)_{Year}$ shows highly skewed pattern (Table 3 and Table 4).

The Citation-per-paper-per-Year (CY) shows the variational pattern for the journals, as $H_0(2)$ is Rejected for the journals (Table 1), but shows constancy for the years, as $H_0(2)$ is accepted for the years (Table 2). The journal JESS possessed highest average CY (1.22) followed by IJBB (0.99), PJP (0.90) and IJP (0.85). The CY remained almost constant over the years with an average of 0.76 (Table 4). The Coefficients of Variation (CV) of CY for journals and years figure 33.3% and 9.2% respectively, where the former is high showing non-constancy and the later is quiet small. The Kurtosis values of 1.01 and -0.903 for $(CY)_{Journal}$ and $(CY)_{Year}$ respectively show skewed pattern of the former one, while the negative Kurtosis value of the later indicates the flat distribution with thin tail revealing constancy. The h-core density (HD) is not constant either for journals or for the years, as $H_0(3)$ is rejected in both cases (Table 1 and Table 2). The values of CV and Kurtosis of HD are 0.108 & -1.172 (Table 3) for the journals, and 0.123 & -0.190 (Table 4) for the years. The low CV values and negative Kurtosis values, however point out the near constancy of HD both for the journals and years, which is also accorded by close proximity of $F_0$ (2.14 (journals) & 3.03 (years)) and $F_C$ (1.86). The Time-normalised h-index (TH) also is not constant for both journals and years as $H_0(4)$ is rejected in both cases. The values of CV and Kurtosis of TH are 0.371 & -0.106 (Table 3) for the journals, and 0.480 & 8.330 (Table 4) for the years. The TH over the years is highly fluctuating, as evident from the high Kurtosis value (8.33, Table 4), while the same for journals is relatively stable as clear from its negative kurtosis value (-0.106, Table 3).

The Uncitedness Factor (UF) for journals is inversely proportional to CY and TH, as $H_0(5)$ and $H_0(6.1)$ for journals are accepted, but it holds no mathematical relationship with HD as $H_0(7)$ is rejected (Table 1). Also, UF is directly proportional to TH over the years, as $H_0(6.2)$ is accepted (Table 2). The Correlation Coefficient (R) between UF and CY is -0.93 (Table 3), which is strong negative correlation. But, the R of UF with TH and HD are -0.46 and 0.52 respectively (Table 3) showing weak negative and weak positive correlations. The R of UF with CY, TH and HD over the years are -0.16, 0.93 and -0.76 respectively (Table 4). The UF has a strong positive correlation with TH over the years, while weak and strong negative correlations with CY and HD respectively.

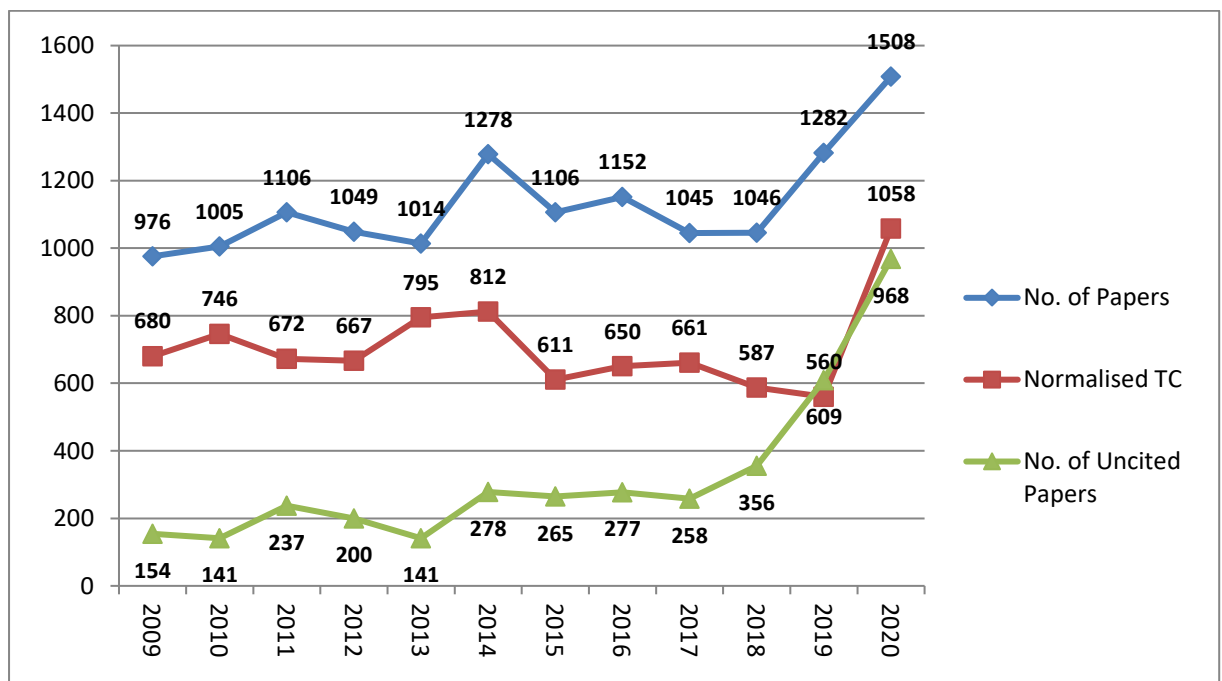

Figure 1: Change in total number of papers (P), time-normalised Total Citation (TC) and number of Uncited papers (U) from 2009 to 2020

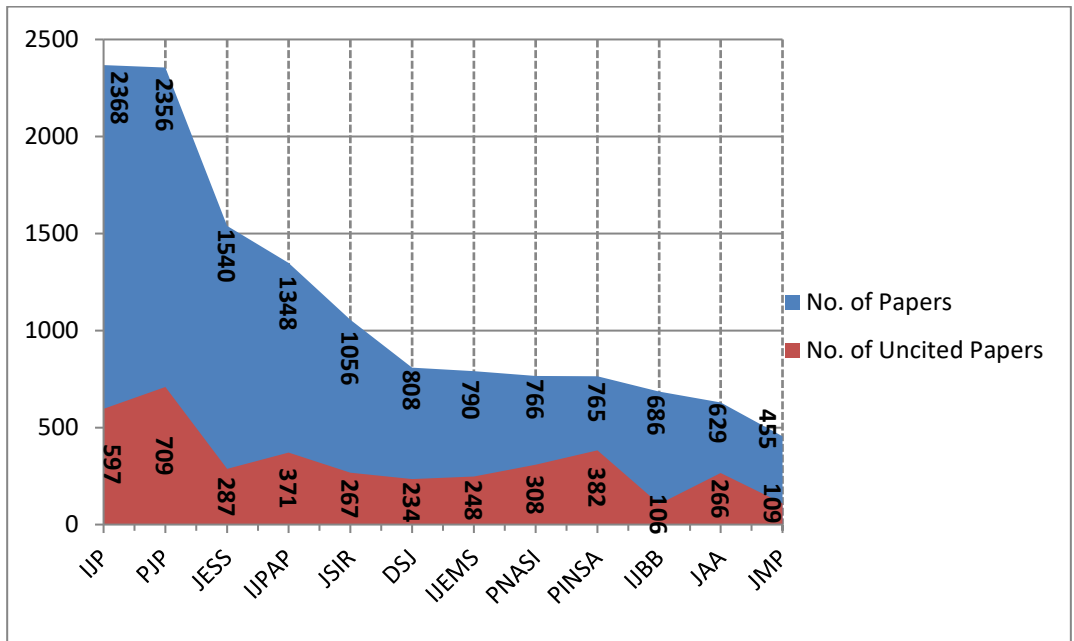

Figure 2: Change in total number of papers (P) and number of Uncited papers (U) for 12 journals

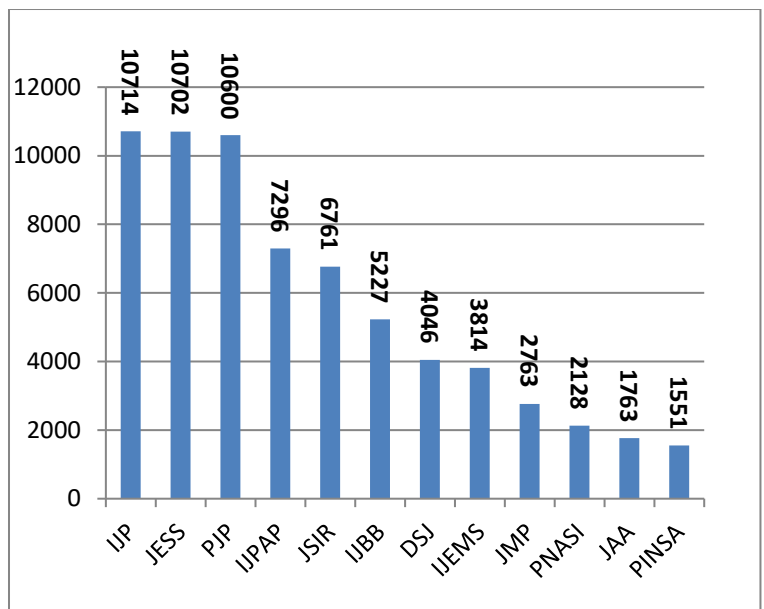

Figure 3: Change in Total number of Citations (TC) for 12 journals

Table 1: Testing of hypothesis for population means of indicators' values for the journals

| Indicators | $F_C$ = 1.86: At 5% level of significance ($\alpha$ = 0.05; df (bg): 11; df (wg): 132) | | | |
|---|---|---|---|---|
| | $F_O$ | P | Observation | Inference: Null Hypothesis ($H_0$) is |
| UF | 2.82 | 0.0025 | $F_C < F_O$; P < $\alpha$ | $H_0(1)$ is Rejected |
| CY | 5.93 | 8.7*10^{-08} | $F_C < F_O$; P < $\alpha$ | $H_0(2)$ is Rejected |
| HD | 2.14 | 0.022 | $F_C < F_O$; P < $\alpha$ | $H_0(3)$ is Rejected |
| TH | 3.61 | 0.0002 | $F_C < F_O$; P < $\alpha$ | $H_0(4)$ is Rejected |
| **UF*CY** | **1.27** | **0.25** | **$F_C > F_O$; P > $\alpha$** | **$H_0(5)$ is Accepted** |
| **UF*TH** | **0.90** | **0.54** | **$F_C > F_O$; P > $\alpha$** | **$H_0(6.1)$ is Accepted** |
| HD/UF | 4.66 | 5.6*10^{-06} | $F_C < F_O$; P < $\alpha$ | $H_0(7)$ is Rejected |

$F_C$ - $F_{Critical}$; $\alpha$ - Level of Significance Value; $F_O$ - $F_{Observed}$ ; P - P-Value; $H_0$ - Null Hypothesis;
df(bg) - Degrees of Freedom (Between groups) = 11; df(wg) - Degrees of Freedom (Within groups) = 132

Table 2: Testing of hypothesis for population means of indicators' values for the years

| Indicators | $F_C = 1.86$: At 5% level of significance ($\alpha = 0.05$; df (bg): 11; df (wg): 132) | | | |
|---|---|---|---|---|
| | $F_O$ | P | Observation | Inference: Null Hypothesis ($H_0$) is |
| UF | 15.96 | $1.7*10^{-19}$ | $F_C < F_O$; $P < \alpha$ | $H_0(1)$ is Rejected |
| **CY** | **0.35** | **0.97** | **$F_C > F_O$; $P > \alpha$** | **$H_0(2)$ is Accepted** |
| HD | 3.03 | 0.001 | $F_C < F_O$; $P < \alpha$ | $H_0(3)$ is Rejected |
| TH | 7.48 | $6.9*10^{-10}$ | $F_C < F_O$; $P < \alpha$ | $H_0(4)$ is Rejected |
| UF*CY | 13.03 | $1.7*10^{-16}$ | $F_C < F_O$; $P < \alpha$ | $H_0(5)$ is Rejected |
| **TH/UF** | **1.11** | **0.361** | **$F_C > F_O$; $P > \alpha$** | **$H_0(6.2)$ is Accepted** |
| HD/UF | 2.75 | 0.003 | $F_C < F_O$; $P < \alpha$ | $H_0(7)$ is Rejected |

Table 3: Statistical parameters of the indicators' values for the journals

| | Mean | Median | Range | Standard Deviation (SD) | Coefficient of Variation (CV) | Kurtosis | Correlation Coefficient (R) |
|---|---|---|---|---|---|---|---|
| UF | 0.28 | 0.28 | 0.33 | 0.093 | 0.333 | 0.049 | --- |
| CY | 0.76 | 0.78 | 0.89 | 0.228 | 0.300 | 1.010 | --- |
| HD | 0.26 | 0.26 | 0.08 | 0.028 | 0.108 | -1.172 | --- |
| TH | 1.86 | 1.69 | 2.00 | 0.691 | 0.371 | -0.106 | --- |
| UF*CY | 0.18 | 0.16 | 0.17 | 0.050 | 0.277 | 2.349 | $R_{UF-CY} = -0.93$ |
| UF*TH | 0.58 | 0.52 | 0.79 | 0.236 | 0.407 | 1.416 | $R_{UF-TH} = -0.46$ |
| HD/UF | 1.79 | 1.49 | 4.52 | 1.206 | 0.674 | 6.567 | $R_{UF-HD} = 0.52$ |

Table 4: Statistical parameters of the indicators' values for the years

| | Mean | Median | Range | Standard Deviation (SD) | Coefficient of Variation (CV) | Kurtosis | Correlation Coefficient (R) |
|---|---|---|---|---|---|---|---|
| UF | 0.28 | 0.22 | 0.52 | 0.162 | 0.577 | 2.766 | --- |
| CY | 0.76 | 0.77 | 0.22 | 0.070 | 0.092 | -0.903 | --- |
| HD | 0.26 | 0.26 | 0.11 | 0.032 | 0.123 | -0.190 | --- |
| TH | 1.86 | 1.58 | 3.29 | 0.890 | 0.480 | 8.330 | --- |
| UF*CY | 0.18 | 0.12 | 0.39 | 0.116 | 0.644 | 4.256 | $R_{UF-CY} = -0.16$ |
| UF*TH | 0.58 | 0.30 | 2.63 | 0.742 | 1.279 | 8.074 | $R_{UF-TH} = 0.93$ |
| TH/UF | 11.82 | 10.16 | 15.65 | 4.681 | 0.396 | -0.52 | |
| HD/UF | 1.80 | 1.89 | 2.90 | 0.993 | 0.552 | -1.240 | $R_{UF-HD} = -0.76$ |

## Findings

1) It is found from the literature review that the percentage of uncited papers in global physics and astronomy research output figured almost 17%, while Indian physics and astronomy research output figured nearly 29% of uncited papers.

2) The variation of UF for the journals indicates that the way in which the number of uncited papers is influenced by the total number of papers is a journal's own attribute. In some journals, large number of papers remain uncited over the years in spite of fair number of regular publications, whereas many journals able to attract citations very quickly. It depends on a journal's indexing status, reputation, get-up and many other factors.

3) It is found that $U = 0.67P * t^{-0.6}$, i.e. number of uncited papers decreases as time passes on. If $t = 1$, then $U = 0.67P$, which indicates the average yearly rate of change of uncited paper is 67% of total number of papers.

4) The Citation-per-paper-per-Year or CY is actually time-normalised average number of citations per paper or time-normalised citation density, which may be interpreted as equivalent to citation potential. It is the indicator of average citability factor of a journal. The temporal constancy of CY signals it's dependency on the subject domain.

5) The factor HD is, though not strictly constant, but still nearly constant as indicated by negative Kurtosis values indicating a steady ratio between h-core and total citations. The h-core citation is thus more or less directly proportional to total citation.

6) The nearly constancy of TH for the journals indicates that h-index is approximately directly proportional to age of publication for Indian physics and astronomy journals.

7) It is found from Table 1, $UF \propto \frac{1}{CY}$ (for the journals), $UF \propto \frac{1}{TH}$ (for the journals) and Hence, Uncitedness Factor (UF) increases with decrease in CY and TH, time-normalised citation density and time-normalised h-index for the journals and vice versa.

8) It is found from Table 2, $UF \propto TH$ (over the years) indicating the change in UF with temporal change in time-normalised h-index (TH) value.

## Conclusion

The Uncitedness Factors of selective venerated Indian physics and astronomy journals are derived here from different aspects. The 12% more uncitedness of Indian physics and astronomy research communication compared to global uncitedness of the same indicates lack of circulation and timely reach of Indian research communication to the pertinent audience. The UF for journals is found to vary inversely with the product of CY and TH, which means an enhance in citation potential (CY) and time-normalised h-index (TH) will reduce the uncitedness factor. The uncitedness factor can be reduced only when the citation becomes scattered or tends to scatter over entire corpus of publications. The scattering nature of citation accelerates coverage of citation, which in turn hastens speed of citation accumulation in accordance with cumulative advantage model. The CY or citation potential and TH or time-normalised h-index depicts the centralised nature of citation distribution. This centralised nature endorses the cumulative advantage model, i.e. success breeds success, that may be viewed here as citation breeds citation, encouraging eventually citation accumulation around highly cited items only. Now, the result of Findings No. 7 suggests that the centralised citation accumulation escalates the citation scattering also, which ultimately reduces uncitedness. Hence, it may be concluded that in spite of higher than global uncitedness, the citation picture of Indian physics journals is dynamic and widespread, but still needs improvement.

## Acknowledgement

This work is executed under the research project entitled *Design and development of comprehensive database and scientometric study of Indian research output in physics and space science since independence* sponsored by Department of Science and Technology, Govt. of India under NSTMIS scheme, (Vide F. No. DST/NSTMIS/05/252/2017-18 dated 11/01/2018).

## References

1. Jacobs G, Cybernetics, homeostasis and a model of disease, *Aerospace Medicine*, 35 (1964) 726-731. PMID: 14210280.

2. Jacobs G, The importance of not being cited, *Journal of American Medical Association*, 217 (5) (1971) 698-699.

3. Tom Wolfe, Radical Chic & Mau-Mauing the Flak Catchers, New York: Farrar, Straus & Giroux, 1970.

4. Garfield, E. (1971). The Road to Scientific Oblivion. *Journal of American Medical Association*, 218(6), 886-887.

5. Garfield, E., & Sher, I. H. (1963). Genetics Citation Index: experimental citation indexes to genetics with special emphasis on human genetics. Philadelphia: Institute for Scientific Information, 1963.

6. Garfield E, When is a negative search result positive, *Current Contents,* (24) (1970), In *Essays of an Information Scientist*, (1) p.117-118.

7. Garfield E, Uncitedness and the identification of dissertation topics, *Current Contents,* (14) (1972), In *Essays of an Information Scientist*, (1) p.291.

8. Ghosh JS and Neufeld ML, Uncitedness of articles in the Journal of the American Chemical Society, *Information Storage and Retrieval,* 10 (11-12) (1974) 365-369.

9. Ghosh JS, Uncitedness of articles in Nature, a multidisciplinary scientific journal, *Information Processing & Management,* 11 (5-7) (1975) 165-169.

10. Lawani S, Some bibliometric correlates of quality in scientific research, *Scientometrics,* 9 (1-2) (1986) 13-25.

11) Stern RE, Uncitedness in the biomedical literature, *Journal of the American society for information science,* 41 (3) (1990) 193-196.

12) Sengupta IN and Henzler R, Citedness and uncitedness of cancer articles, *Scientometrics,* 22 (2) (1991) 283-296.

13) Szava-Kovats E, Non-Indexed Eponymal Citedness (NIEC): first fact-finding examination of a phenomenon of scientific literature, *Journal of Information Science,* 20 (1) (1994) 55-70.

14) Hamilton DP, Research papers: who's uncited now? *Science,* 251 (1991) 25.

15) Hamilton DP, Publishing by -- and for? -- the numbers, *Science,* 250 (1990) 1331-2.

16) Pendlebury DA, Science, citation and funding, *Science*, 251 (1991) 1410-1.

17) Garfield E, I had a dream... about uncitedness, *The Scientist,* 12 (14) (1998) 10.

18) Schwartz CA, The rise and fall of uncitedness, *College & Research Libraries,* 58 (1) (1997) 19-29.

19) Van Dalen HP and Kène H, Demographers and their journals: who remains uncited after ten years? *Population and Development review,* 30 (3) (2004) 489-506.

20) Small H, On the shoulders of Robert Merton: towards a normative theory of citation, *Scientometrics,* 60 (1) (2004) 71-79.

21) Leeuwen V, Thed N and Moed HF, Characteristics of journal impact factors: the effects of uncitedness and citation distribution on the understanding of journal impact factors, *Scientometrics,* 63 (2) (2005) 357-371.

22) Van Dalen, Hendrik P and Henkens K, Signals in science-on the importance of signalling in gaining attention in science, *Scientometrics,* 64 (2) (2005) 209-233.

23) Egghe L, The mathematical relation between the impact factor and the uncitedness factor, *Scientometrics,* 76 (1) (2008) 117-123.

24) Onyancha O, A citation analysis of sub-Saharan African library and information science journals using Google Scholar, *African Journal of Library Archives and Information Science*, 19 (2) (2009) 101-116.

25) Wallace ML, Vincent L and Yves G, Modeling a century of citation distributions, *Journal of Informetrics,* 3 (4) (2009) 296-303.

26) Egghe L, The distribution of the uncitedness factor and its functional relation with the impact factor, *Scientometrics,* 83 (3) (2010) 689-695.

27) Egghe L, Guns R and Rousseau R, Thoughts on uncitedness: Nobel laureates and fields medallists as case studies, *Journal of the American Society for Information Science and Technology*, 62 (8) (2011) 1637-1644.

28) Hsu JW and Ding-wei H, A scaling between impact factor and uncitedness, *Physica A: Statistical Mechanics and its Applications,* 391 (5) (2012) 2129-2134.

29) Burrell QL, Alternative thoughts on uncitedness, *Journal of the American Society for Information Science and Technology,* 63 (7) (2012) 1466-1470.

30) Burrell QL, A stochastic approach to the relation between the impact factor and the uncitedness factor, *Journal of Informetrics,* 7 (3) (2013) 676-682.

31) Heneberg P, Supposedly uncited articles of Nobel laureates and Fields medallists can be prevalently attributed to the errors of omission and commission, *Journal of the American Society for Information Science and Technology,* 64 (3) (2013) 448-454.

32) Egghe L, The functional relation between the impact factor and the uncitedness factor revisited, *Journal of Informetrics,* 7 (1) (2013) 183-189.

33) Law R, Hee Andy L and Norman A, Which journal articles are uncited? The case of the *Asia Pacific Journal of Tourism Research* and *the Journal of Travel and Tourism Marketing*, *Asia Pacific Journal of Tourism Research*, 18 (6) (2013) 661-684.

34) Garg KC and S Kumar, Uncitedness of Indian scientific output, *Current Science*, 107 (6) (2014) 965-970.

35) Lou W and He J, Does author affiliation reputation affect uncitedness? *Proceedings of the Association for Information Science and Technology*, 52 (1) (2015) 1-4.

36) Arsenault C and Vincent L, Is paper uncitedness a function of the alphabet?." *International Society for Scientometrics and Informetrics Archive,* (2015), (Available at https://www.issi-society.org/proceedings/issi_2015).

37) Liang L, Zhong Z and Rousseau R, Uncited papers, uncited authors and uncited topics: A case study in library and information science, *Journal of Informetrics*, 9 (1) (2015) 50-58.

38) Gopalakrishnan S, Bathrinarayanan AL and Tamizhchelvan M, Uncited publications in MEMS literature: A bibliometric study, *DESIDOC Journal of Library & Information Technology*, 35 (2) (2015) 113-123.

39) Elango B, Uncitedness in scientific publications: a case study of tribology research, *SRELS Journal of Information Management* 53.4 (2016): 293-296.

40) Zewen H and Yishan W, A probe into causes of non-citation based on survey data, *Social Science Information*, 57 (1) (2018) 139-151.

41) Zewen H, Yishan W and Jianjun S, A quantitative analysis of determinants of non-citation using a panel data model, *Scientometrics*, 116 (2) (2018) 843-861.

42) Zewen H, Yishan W and Jianjun S, A survey-based structural equation model analysis on influencing factors of non-citation, *Current Science*, 114 (11) (2018) 2302.

43) Yeung, AWK, Higher impact factor of neuroimaging journals is associated with larger number of articles published and smaller percentage of uncited articles, *Frontiers in human neuroscience*, 12 (2019) 523.

44) Nowroozzadeh MH and Salehi-Marzijarani M, Uncitedness in the top general medical journals, *Journal of general internal medicine*, 34 (12) (2019) 2695-2696.

45) Nicolaisen J and Tove FF, Zero impact: a large-scale study of uncitedness, *Scientometrics,* 119 (2) (2019) 1227-1254.

46) Baruch Y, Fabian H and Abdul Rahman A, Are half of the published papers in top-management-journals never ever cited? Refuting the myth and examining the reason for its creation, *Studies in Higher Education* (2020) 1-16.

47) Dorta-González P, Rafael SV and María Isabel DG, Open access effect on uncitedness: a large-scale study controlling by discipline, source type and visibility, *Scientometrics*, 124 (3) (2020) 2619-2644.

48) Lloyd M and Ordorika I, International university rankings as cultural imperialism: implications for the global south, In Global university rankings and the politics of knowledge, edited by Stack M (University of Toronto Press; Toronto), 2021, p. 25-49.

49) Price DJS, Networks of scientific papers, *Science*, 149 (1965) 510-515.

50) Price DJS, A general theory of bibliometric and other cumulative advantage processes, *Journal of the American society for Information science*, 27 (5) (1976) 292-306.

51) Merton RK, The Matthew effect in science, II: cumulative advantage and the symbolism of intellectual property, *Isis: A Journal of the History of Science Society*, 79 (4) (1988) 606-623.

52) Vitanov NK, Science dynamics and research production: indicators, indexes, statistical laws and mathematical models. Cham: Springer, 2016, p.63.

# Appendix

### Table A1: Uncitedness Factor (UF)

|         | 2009 | 2010 | 2011 | 2012 | 2013 | 2014 | 2015 | 2016 | 2017 | 2018 | 2019 | 2020 | Mean(J) |
|---------|------|------|------|------|------|------|------|------|------|------|------|------|---------|
| DSJ     | 0.11 | 0.20 | 0.21 | 0.12 | 0.22 | 0.11 | 0.11 | 0.19 | 0.35 | 0.28 | 0.53 | 0.73 | **0.26** |
| IJBB    | 0.07 | 0.00 | 0.07 | 0.08 | 0.04 | 0.04 | 0.28 | 0.25 | 0.31 | 0.24 | 0.27 | 0.37 | **0.17** |
| IJEMS   | 0.13 | 0.08 | 0.08 | 0.00 | 0.11 | 0.10 | 0.16 | 0.26 | 0.33 | 0.64 | 0.74 | 0.87 | **0.29** |
| IJP     | 0.15 | 0.13 | 0.27 | 0.05 | 0.08 | 0.12 | 0.19 | 0.15 | 0.20 | 0.28 | 0.49 | 0.54 | **0.22** |
| IJPAP   | 0.13 | 0.08 | 0.11 | 0.23 | 0.11 | 0.18 | 0.26 | 0.25 | 0.33 | 0.40 | 0.75 | 0.84 | **0.30** |
| JAA     | 0.07 | 0.35 | 0.42 | 0.37 | 0.33 | 0.58 | 0.11 | 0.26 | 0.29 | 0.49 | 0.56 | 0.76 | **0.38** |
| JESS    | 0.07 | 0.03 | 0.02 | 0.02 | 0.04 | 0.04 | 0.07 | 0.06 | 0.13 | 0.21 | 0.43 | 0.54 | **0.14** |
| JMP     | 0.10 | 0.13 | 0.09 | 0.14 | 0.11 | 0.12 | 0.13 | 0.34 | 0.20 | 0.40 | 0.50 | 0.79 | **0.25** |
| JSIR    | 0.11 | 0.08 | 0.10 | 0.12 | 0.16 | 0.12 | 0.26 | 0.20 | 0.07 | 0.25 | 0.35 | 0.81 | **0.22** |
| PINSA   | 0.29 | 0.39 | 0.45 | 0.37 | 0.33 | 0.34 | 0.55 | 0.45 | 0.35 | 0.63 | 0.67 | 0.79 | **0.47** |
| PJP     | 0.21 | 0.20 | 0.23 | 0.40 | 0.22 | 0.37 | 0.31 | 0.30 | 0.25 | 0.21 | 0.34 | 0.56 | **0.30** |
| PNASI   | 0.53 | 0.43 | 0.63 | 0.18 | 0.23 | 0.28 | 0.26 | 0.20 | 0.18 | 0.41 | 0.49 | 0.62 | **0.37** |
| Mean(Y) | **0.16** | **0.17** | **0.22** | **0.17** | **0.17** | **0.20** | **0.22** | **0.24** | **0.25** | **0.37** | **0.51** | **0.68** | |

### Table A2: Citation-per-Paper-per-Year (CY)

|         | 2009 | 2010 | 2011 | 2012 | 2013 | 2014 | 2015 | 2016 | 2017 | 2018 | 2019 | 2020 | Mean(J) |
|---------|------|------|------|------|------|------|------|------|------|------|------|------|---------|
| DSJ     | 1.07 | 0.98 | 0.77 | 0.85 | 0.64 | 0.84 | 0.67 | 0.98 | 0.62 | 0.89 | 0.48 | 0.42 | **0.77** |
| IJBB    | 1.49 | 1.42 | 1.22 | 1.07 | 1.18 | 0.89 | 0.54 | 0.33 | 0.45 | 0.74 | 0.83 | 1.68 | **0.99** |
| IJEMS   | 0.55 | 0.98 | 1.06 | 0.93 | 0.99 | 0.72 | 0.70 | 0.83 | 0.52 | 0.34 | 0.19 | 0.23 | **0.67** |
| IJP     | 0.48 | 0.59 | 0.56 | 0.77 | 0.98 | 0.92 | 0.73 | 0.88 | 0.96 | 1.13 | 0.72 | 1.54 | **0.85** |
| IJPAP   | 0.69 | 1.03 | 0.82 | 0.68 | 0.76 | 0.80 | 0.69 | 0.76 | 0.64 | 0.66 | 0.31 | 0.26 | **0.68** |
| JAA     | 0.47 | 0.23 | 0.37 | 0.39 | 0.73 | 0.20 | 0.69 | 0.49 | 1.15 | 0.48 | 0.56 | 0.45 | **0.52** |
| JESS    | 1.41 | 1.53 | 1.37 | 1.54 | 1.55 | 1.19 | 1.35 | 1.12 | 1.20 | 0.82 | 0.57 | 1.02 | **1.22** |
| JMP     | 0.87 | 1.49 | 0.83 | 0.90 | 1.00 | 0.94 | 0.73 | 0.73 | 0.89 | 0.40 | 0.44 | 0.25 | **0.79** |
| JSIR    | 1.46 | 1.28 | 0.81 | 0.72 | 0.77 | 0.86 | 0.51 | 0.54 | 0.93 | 0.76 | 0.68 | 0.42 | **0.81** |
| PINSA   | 0.20 | 0.30 | 0.12 | 0.42 | 0.17 | 0.75 | 0.28 | 0.39 | 0.45 | 0.24 | 0.25 | 0.39 | **0.33** |
| PJP     | 0.59 | 0.52 | 0.58 | 0.55 | 0.87 | 0.39 | 0.65 | 0.76 | 1.11 | 1.23 | 1.47 | 2.12 | **0.90** |
| PNASI   | 0.11 | 0.18 | 0.10 | 0.51 | 0.25 | 0.74 | 0.62 | 0.89 | 0.89 | 0.80 | 1.41 | 1.16 | **0.64** |
| Mean(Y) | **0.78** | **0.88** | **0.72** | **0.78** | **0.82** | **0.77** | **0.68** | **0.72** | **0.82** | **0.71** | **0.66** | **0.83** | |

### Table A3: h-Core Density (HD)

|         | 2009 | 2010 | 2011 | 2012 | 2013 | 2014 | 2015 | 2016 | 2017 | 2018 | 2019 | 2020 | Mean(J) |
|---------|------|------|------|------|------|------|------|------|------|------|------|------|---------|
| DSJ     | 0.21 | 0.31 | 0.30 | 0.31 | 0.33 | 0.27 | 0.29 | 0.28 | 0.24 | 0.20 | 0.23 | 0.23 | **0.27** |
| IJBB    | 0.31 | 0.32 | 0.27 | 0.29 | 0.32 | 0.27 | 0.18 | 0.23 | 0.28 | 0.23 | 0.16 | 0.18 | **0.25** |
| IJEMS   | 0.28 | 0.33 | 0.35 | 0.39 | 0.26 | 0.27 | 0.35 | 0.33 | 0.20 | 0.26 | 0.25 | 0.35 | **0.30** |
| IJP     | 0.21 | 0.25 | 0.25 | 0.19 | 0.19 | 0.18 | 0.25 | 0.23 | 0.28 | 0.21 | 0.18 | 0.23 | **0.22** |

| | | | | | | | | | | | | |
|---|---|---|---|---|---|---|---|---|---|---|---|---|
| IJPAP | 0.25 | 0.25 | 0.26 | 0.28 | 0.26 | 0.26 | 0.23 | 0.21 | 0.25 | 0.21 | 0.13 | 0.21 | **0.23** |
| JAA | 0.46 | 0.21 | 0.33 | 0.37 | 0.28 | 0.22 | 0.25 | 0.25 | 0.37 | 0.23 | 0.29 | 0.31 | **0.30** |
| JESS | 0.33 | 0.26 | 0.30 | 0.30 | 0.24 | 0.19 | 0.25 | 0.21 | 0.22 | 0.15 | 0.09 | 0.19 | **0.23** |
| JMP | 0.29 | 0.16 | 0.34 | 0.21 | 0.29 | 0.37 | 0.28 | 0.30 | 0.22 | 0.33 | 0.32 | 0.14 | **0.27** |
| JSIR | 0.24 | 0.22 | 0.26 | 0.29 | 0.20 | 0.20 | 0.17 | 0.24 | 0.48 | 0.29 | 0.12 | 0.15 | **0.24** |
| PINSA | 0.49 | 0.21 | 0.23 | 0.34 | 0.40 | 0.31 | 0.27 | 0.26 | 0.25 | 0.18 | 0.17 | 0.24 | **0.28** |
| PJP | 0.23 | 0.23 | 0.26 | 0.28 | 0.27 | 0.14 | 0.19 | 0.16 | 0.21 | 0.19 | 0.20 | 0.29 | **0.22** |
| PNASI | 0.25 | 0.29 | 0.41 | 0.28 | 0.15 | 0.41 | 0.28 | 0.28 | 0.27 | 0.21 | 0.16 | 0.14 | **0.26** |
| Mean(Y) | **0.29** | **0.25** | **0.30** | **0.29** | **0.27** | **0.26** | **0.25** | **0.25** | **0.27** | **0.22** | **0.19** | **0.22** | |

Table A4: Time-Normalised h-Index (TH)

| | 2009 | 2010 | 2011 | 2012 | 2013 | 2014 | 2015 | 2016 | 2017 | 2018 | 2019 | 2020 | Mean(J) |
|---|---|---|---|---|---|---|---|---|---|---|---|---|---|
| DSJ | 1.08 | 1.18 | 1.20 | 1.22 | 1.38 | 1.43 | 1.33 | 2.00 | 1.75 | 2.00 | 2.00 | 3.00 | **1.63** |
| IJBB | 1.67 | 1.55 | 1.40 | 1.44 | 1.88 | 1.57 | 0.83 | 0.60 | 1.00 | 1.67 | 2.00 | 5.00 | **1.72** |
| IJEMS | 0.92 | 1.36 | 1.40 | 1.33 | 1.50 | 1.57 | 1.83 | 1.60 | 1.25 | 1.33 | 1.00 | 3.00 | **1.51** |
| IJP | 1.17 | 1.55 | 1.70 | 1.67 | 2.13 | 2.14 | 2.17 | 2.60 | 3.50 | 3.67 | 4.00 | 11.0 | **3.11** |
| IJPAP | 1.42 | 1.82 | 1.60 | 1.78 | 1.75 | 1.71 | 1.67 | 1.80 | 2.00 | 2.00 | 1.50 | 2.00 | **1.75** |
| JAA | 0.50 | 0.27 | 1.20 | 0.56 | 0.88 | 0.86 | 1.17 | 1.00 | 2.75 | 1.67 | 2.00 | 2.00 | **1.24** |
| JESS | 1.50 | 1.55 | 1.90 | 2.33 | 2.38 | 2.14 | 2.67 | 2.60 | 3.00 | 2.33 | 2.50 | 6.00 | **2.57** |
| JMP | 0.92 | 0.91 | 1.00 | 0.89 | 1.13 | 1.43 | 1.17 | 1.40 | 1.50 | 1.33 | 1.50 | 1.00 | **1.18** |
| JSIR | 1.83 | 1.64 | 1.50 | 1.33 | 1.25 | 1.57 | 1.17 | 1.40 | 1.25 | 2.00 | 2.00 | 3.00 | **1.66** |
| PINSA | 0.42 | 0.36 | 0.30 | 1.00 | 0.50 | 1.71 | 1.17 | 1.60 | 1.25 | 1.00 | 1.00 | 3.00 | **1.11** |
| PJP | 1.42 | 1.55 | 1.70 | 2.00 | 2.25 | 1.29 | 2.00 | 2.40 | 3.25 | 3.67 | 5.50 | 10.0 | **3.08** |
| PNASI | 1.67 | 1.55 | 1.40 | 1.44 | 1.88 | 1.57 | 0.83 | 0.60 | 1.00 | 1.67 | 2.00 | 5.00 | **1.72** |
| Mean(Y) | **1.21** | **1.27** | **1.36** | **1.42** | **1.57** | **1.58** | **1.50** | **1.63** | **1.96** | **2.03** | **2.25** | **4.50** | |

Table A5: UF*CY

| | 2009 | 2010 | 2011 | 2012 | 2013 | 2014 | 2015 | 2016 | 2017 | 2018 | 2019 | 2020 | Mean(J) |
|---|---|---|---|---|---|---|---|---|---|---|---|---|---|
| DSJ | 0.12 | 0.20 | 0.16 | 0.10 | 0.14 | 0.09 | 0.07 | 0.19 | 0.22 | 0.25 | 0.25 | 0.31 | **0.17** |
| IJBB | 0.10 | 0.00 | 0.08 | 0.09 | 0.05 | 0.04 | 0.15 | 0.08 | 0.14 | 0.18 | 0.22 | 0.62 | **0.15** |
| IJEMS | 0.07 | 0.08 | 0.08 | 0.00 | 0.11 | 0.07 | 0.11 | 0.21 | 0.17 | 0.22 | 0.14 | 0.20 | **0.12** |
| IJP | 0.07 | 0.08 | 0.15 | 0.04 | 0.08 | 0.11 | 0.14 | 0.13 | 0.20 | 0.32 | 0.35 | 0.83 | **0.21** |
| IJPAP | 0.09 | 0.08 | 0.09 | 0.15 | 0.09 | 0.14 | 0.18 | 0.19 | 0.21 | 0.26 | 0.23 | 0.22 | **0.16** |
| JAA | 0.03 | 0.08 | 0.16 | 0.14 | 0.24 | 0.11 | 0.07 | 0.13 | 0.34 | 0.24 | 0.31 | 0.34 | **0.18** |
| JESS | 0.10 | 0.05 | 0.03 | 0.03 | 0.06 | 0.05 | 0.10 | 0.07 | 0.16 | 0.18 | 0.24 | 0.55 | **0.13** |
| JMP | 0.09 | 0.20 | 0.07 | 0.12 | 0.11 | 0.11 | 0.09 | 0.25 | 0.18 | 0.16 | 0.22 | 0.20 | **0.15** |
| JSIR | 0.17 | 0.10 | 0.08 | 0.09 | 0.12 | 0.10 | 0.13 | 0.11 | 0.07 | 0.19 | 0.24 | 0.34 | **0.14** |
| PINSA | 0.06 | 0.12 | 0.05 | 0.15 | 0.06 | 0.25 | 0.15 | 0.18 | 0.16 | 0.15 | 0.17 | 0.30 | **0.15** |
| PJP | 0.12 | 0.10 | 0.13 | 0.12 | 0.22 | 0.19 | 0.14 | 0.21 | 0.23 | 0.28 | 0.50 | 1.19 | **0.30** |
| PNASI | 0.06 | 0.08 | 0.06 | 0.09 | 0.06 | 0.21 | 0.16 | 0.18 | 0.16 | 0.33 | 0.69 | 0.72 | **0.23** |
| Mean(Y) | **0.09** | **0.10** | **0.10** | **0.10** | **0.11** | **0.12** | **0.13** | **0.16** | **0.19** | **0.23** | **0.30** | **0.48** | |

Table A6: UF*TH

| | 2009 | 2010 | 2011 | 2012 | 2013 | 2014 | 2015 | 2016 | 2017 | 2018 | 2019 | 2020 | Mean(J) |
|---|---|---|---|---|---|---|---|---|---|---|---|---|---|
| DSJ | 0.12 | 0.24 | 0.25 | 0.14 | 0.31 | 0.16 | 0.15 | 0.38 | 0.61 | 0.57 | 1.05 | 2.19 | **0.51** |
| IJBB | 0.12 | 0.00 | 0.09 | 0.12 | 0.08 | 0.07 | 0.23 | 0.15 | 0.31 | 0.40 | 0.53 | 1.83 | **0.33** |
| IJEMS | 0.12 | 0.11 | 0.11 | 0.00 | 0.17 | 0.16 | 0.29 | 0.41 | 0.42 | 0.86 | 0.74 | 2.61 | **0.50** |
| IJP | 0.18 | 0.20 | 0.46 | 0.08 | 0.18 | 0.25 | 0.41 | 0.40 | 0.71 | 1.04 | 1.95 | 5.94 | **0.98** |
| IJPAP | 0.18 | 0.14 | 0.17 | 0.40 | 0.20 | 0.30 | 0.43 | 0.46 | 0.67 | 0.79 | 1.12 | 1.67 | **0.54** |
| JAA | 0.04 | 0.10 | 0.51 | 0.20 | 0.29 | 0.50 | 0.12 | 0.26 | 0.80 | 0.82 | 1.12 | 1.52 | **0.52** |
| **JESS** | 0.10 | 0.05 | 0.04 | 0.04 | 0.10 | 0.09 | 0.19 | 0.16 | 0.40 | 0.50 | 1.06 | 3.24 | **0.50** |
| **JMP** | 0.09 | 0.12 | 0.09 | 0.12 | 0.13 | 0.17 | 0.15 | 0.48 | 0.30 | 0.53 | 0.75 | 0.79 | **0.31** |
| **JSIR** | 0.21 | 0.13 | 0.16 | 0.16 | 0.20 | 0.19 | 0.30 | 0.28 | 0.09 | 0.51 | 0.69 | 2.43 | **0.44** |
| **PINSA** | 0.12 | 0.14 | 0.14 | 0.37 | 0.17 | 0.58 | 0.64 | 0.73 | 0.43 | 0.63 | 0.67 | 2.36 | **0.58** |
| **PJP** | 0.29 | 0.31 | 0.39 | 0.80 | 0.50 | 0.48 | 0.63 | 0.73 | 0.81 | 0.75 | 1.86 | 5.61 | **1.10** |
| **PNASI** | 0.18 | 0.19 | 0.25 | 0.14 | 0.06 | 0.48 | 0.35 | 0.37 | 0.41 | 0.83 | 1.46 | 3.11 | **0.65** |
| **Mean(Y)** | **0.15** | **0.14** | **0.22** | **0.21** | **0.20** | **0.29** | **0.32** | **0.40** | **0.50** | **0.69** | **1.09** | **2.77** | |

Table A7: HD/UF

| | 2009 | 2010 | 2011 | 2012 | 2013 | 2014 | 2015 | 2016 | 2017 | 2018 | 2019 | 2020 | Mean(J) |
|---|---|---|---|---|---|---|---|---|---|---|---|---|---|
| DSJ | 1.89 | 1.57 | 1.44 | 2.65 | 1.48 | 2.43 | 2.64 | 1.46 | 0.68 | 0.71 | 0.43 | 0.32 | **1.47** |
| IJBB | 4.46 | 0 | 4.02 | 3.52 | 7.96 | 6.46 | 0.64 | 0.92 | 0.88 | 0.94 | 0.61 | 0.50 | **2.58** |
| IJEMS | 2.05 | 4.19 | 4.64 | 0 | 2.27 | 2.66 | 2.23 | 1.29 | 0.60 | 0.41 | 0.34 | 0.40 | **1.92** |
| IJP | 1.42 | 1.93 | 0.94 | 4.07 | 2.30 | 1.59 | 1.34 | 1.48 | 1.37 | 0.73 | 0.36 | 0.43 | **1.50** |
| IJPAP | 1.93 | 3.22 | 2.41 | 1.23 | 2.30 | 1.50 | 0.90 | 0.82 | 0.75 | 0.53 | 0.18 | 0.25 | **1.34** |
| JAA | 6.38 | 0.59 | 0.79 | 1.01 | 0.84 | 0.38 | 2.36 | 0.94 | 1.25 | 0.46 | 0.51 | 0.41 | **1.33** |
| JESS | 4.80 | 8.61 | 13.19 | 15.87 | 5.81 | 4.51 | 3.52 | 3.36 | 1.66 | 0.71 | 0.22 | 0.35 | **5.22** |
| JMP | 2.89 | 1.22 | 4.00 | 1.57 | 2.54 | 3.04 | 2.25 | 0.89 | 1.12 | 0.83 | 0.64 | 0.18 | **1.77** |
| JSIR | 2.12 | 2.88 | 2.53 | 2.47 | 1.26 | 1.68 | 0.67 | 1.20 | 6.73 | 1.12 | 0.36 | 0.19 | **1.93** |
| PINSA | 1.72 | 0.53 | 0.51 | 0.94 | 1.20 | 0.91 | 0.50 | 0.57 | 0.72 | 0.28 | 0.26 | 0.30 | **0.70** |
| PJP | 1.10 | 1.15 | 1.16 | 0.69 | 1.20 | 0.38 | 0.61 | 0.53 | 0.83 | 0.94 | 0.61 | 0.51 | **0.81** |
| PNASI | 0.47 | 0.69 | 0.66 | 1.53 | 0.67 | 1.46 | 1.07 | 1.40 | 1.51 | 0.52 | 0.32 | 0.22 | **0.88** |
| **Mean(Y)** | **2.60** | **2.21** | **3.02** | **3.23** | **2.49** | **2.25** | **1.56** | **1.24** | **1.51** | **0.68** | **0.40** | **0.34** | |

Table A8: The Primary Data for Calculation of Indicators Presented in Table 1 to Table 4

| Year | 2009 | 2010 | 2011 | 2012 | 2013 | 2014 | 2015 | 2016 | 2017 | 2018 | 2019 | 2020 | Total |
|---|---|---|---|---|---|---|---|---|---|---|---|---|---|
| **Defence Science Journal** | | | | | | | | | | | | | |
| No. of papers (P) | 64 | 50 | 63 | 52 | 72 | 62 | 54 | 74 | 83 | 67 | 74 | 93 | 808 |
| No. of uncited papers (U) | 7 | 10 | 13 | 6 | 16 | 7 | 6 | 14 | 29 | 19 | 39 | 68 | 234 |
| Total citation (TC) | 818 | 539 | 485 | 396 | 368 | 365 | 218 | 362 | 207 | 178 | 71 | 39 | 4046 |
| h-index (h) | 13 | 13 | 12 | 11 | 11 | 10 | 8 | 10 | 7 | 6 | 4 | 3 | |
| **Indian Journal of Biochemistry and Biophysics** | | | | | | | | | | | | | |
| No. of papers (P) | 72 | 58 | 59 | 61 | 74 | 71 | 43 | 24 | 32 | 50 | 60 | 82 | 686 |
| No. of uncited papers (U) | 5 | 0 | 4 | 5 | 3 | 3 | 12 | 6 | 10 | 12 | 16 | 30 | 106 |
| Total citation (TC) | 1291 | 908 | 719 | 585 | 697 | 443 | 139 | 39 | 58 | 111 | 99 | 138 | 5227 |
| h-index (h) | 20 | 17 | 14 | 13 | 15 | 11 | 5 | 3 | 4 | 5 | 4 | 5 | |
| **Indian Journal of Engineering and Materials Sciences** | | | | | | | | | | | | | |
| No. of papers (P) | 67 | 63 | 53 | 44 | 71 | 88 | 82 | 47 | 60 | 59 | 42 | 114 | 790 |
| No. of uncited papers (U) | 9 | 5 | 4 | 0 | 8 | 9 | 13 | 12 | 20 | 38 | 31 | 99 | 248 |
| Total citation (TC) | 439 | 676 | 560 | 368 | 562 | 445 | 342 | 195 | 124 | 61 | 16 | 26 | 3814 |
| h-index (h) | 11 | 15 | 14 | 12 | 12 | 11 | 11 | 8 | 5 | 4 | 2 | 3 | |
| **Indian Journal of Physics** | | | | | | | | | | | | | |
| No. of papers (P) | 160 | 181 | 204 | 169 | 189 | 190 | 154 | 171 | 183 | 172 | 254 | 341 | 2368 |
| No. of uncited papers (U) | 24 | 23 | 55 | 8 | 16 | 22 | 29 | 26 | 37 | 49 | 124 | 184 | 597 |
| Total citation (TC) | 920 | 1178 | 1146 | 1167 | 1484 | 1219 | 670 | 750 | 706 | 584 | 364 | 526 | 10714 |
| h-index (h) | 14 | 17 | 17 | 15 | 17 | 15 | 13 | 13 | 14 | 11 | 8 | 11 | |
| **Indian Journal of Pure and Applied Physics** | | | | | | | | | | | | | |
| No. of papers (P) | 138 | 142 | 121 | 150 | 124 | 97 | 105 | 102 | 99 | 86 | 111 | 73 | 1348 |
| No. of uncited papers (U) | 18 | 11 | 13 | 34 | 14 | 17 | 27 | 26 | 33 | 34 | 83 | 61 | 371 |
| Total citation (TC) | 1150 | 1605 | 989 | 917 | 755 | 546 | 433 | 388 | 255 | 171 | 68 | 19 | 7296 |

| h-index (h) | 17 | 20 | 16 | 16 | 14 | 12 | 10 | 9 | 8 | 6 | 3 | 2 | |
|---|---|---|---|---|---|---|---|---|---|---|---|---|---|

**Journal of Astrophysics and Astronomy**

| | | | | | | | | | | | | | |
|---|---|---|---|---|---|---|---|---|---|---|---|---|---|
| No. of papers (P) | 14 | 17 | 116 | 19 | 30 | 118 | 47 | 42 | 72 | 75 | 50 | 29 | 629 |
| No. of uncited papers (U) | 1 | 6 | 49 | 7 | 10 | 69 | 5 | 11 | 21 | 37 | 28 | 22 | 266 |
| Total citation (TC) | 79 | 43 | 431 | 67 | 175 | 162 | 195 | 102 | 331 | 109 | 56 | 13 | 1763 |
| h-index (h) | 6 | 3 | 12 | 5 | 7 | 6 | 7 | 5 | 11 | 5 | 4 | 2 | |

**Journal of Earth System Science**

| | | | | | | | | | | | | | |
|---|---|---|---|---|---|---|---|---|---|---|---|---|---|
| No. of papers (P) | 59 | 66 | 88 | 106 | 121 | 140 | 127 | 143 | 135 | 131 | 235 | 189 | 1540 |
| No. of uncited papers (U) | 4 | 2 | 2 | 2 | 5 | 6 | 9 | 9 | 18 | 28 | 100 | 102 | 287 |
| Total citation (TC) | 996 | 1108 | 1204 | 1473 | 1503 | 1163 | 1026 | 799 | 649 | 322 | 266 | 193 | 10702 |
| h-index (h) | 18 | 17 | 19 | 21 | 19 | 15 | 16 | 13 | 12 | 7 | 5 | 6 | |

**Journal of Medical Physics**

| | | | | | | | | | | | | | |
|---|---|---|---|---|---|---|---|---|---|---|---|---|---|
| No. of papers (P) | 40 | 38 | 35 | 37 | 35 | 41 | 40 | 44 | 45 | 40 | 32 | 28 | 455 |
| No. of uncited papers (U) | 4 | 5 | 3 | 5 | 4 | 5 | 5 | 15 | 9 | 16 | 16 | 22 | 109 |
| Total citation (TC) | 418 | 624 | 292 | 301 | 279 | 270 | 174 | 161 | 161 | 48 | 28 | 7 | 2763 |
| h-index (h) | 11 | 10 | 10 | 8 | 9 | 10 | 7 | 7 | 6 | 4 | 3 | 1 | |

**Journal of Scientific and Industrial Research**

| | | | | | | | | | | | | | |
|---|---|---|---|---|---|---|---|---|---|---|---|---|---|
| No. of papers (P) | 114 | 103 | 106 | 76 | 81 | 101 | 94 | 76 | 14 | 55 | 95 | 141 | 1056 |
| No. of uncited papers (U) | 13 | 8 | 11 | 9 | 13 | 12 | 24 | 15 | 1 | 14 | 33 | 114 | 267 |
| Total citation (TC) | 2002 | 1447 | 857 | 492 | 496 | 607 | 287 | 207 | 52 | 126 | 129 | 59 | 6761 |
| h-index (h) | 22 | 18 | 15 | 12 | 10 | 11 | 7 | 7 | 5 | 6 | 4 | 3 | |

**Proceedings of the Indian National Science Academy**

| | | | | | | | | | | | | | |
|---|---|---|---|---|---|---|---|---|---|---|---|---|---|
| No. of papers (P) | 21 | 23 | 33 | 63 | 30 | 89 | 108 | 128 | 55 | 71 | 46 | 98 | 765 |
| No. of uncited papers (U) | 6 | 9 | 15 | 23 | 10 | 30 | 59 | 58 | 19 | 45 | 31 | 77 | 382 |
| Total citation (TC) | 51 | 77 | 39 | 236 | 40 | 467 | 180 | 249 | 100 | 51 | 23 | 38 | 1551 |
| h-index (h) | 5 | 4 | 3 | 9 | 4 | 12 | 7 | 8 | 5 | 3 | 2 | 3 | |

**Pramana - Journal of Physics**

| | | | | | | | | | | | | | |
|---|---|---|---|---|---|---|---|---|---|---|---|---|---|
| No. of papers (P) | 178 | 222 | 188 | 234 | 174 | 213 | 191 | 237 | 184 | 170 | 201 | 164 | 2356 |
| No. of uncited papers (U) | 37 | 44 | 43 | 94 | 39 | 79 | 60 | 72 | 46 | 35 | 68 | 92 | 709 |
| Total citation (TC) | 1260 | 1268 | 1093 | 1161 | 1205 | 582 | 748 | 899 | 818 | 628 | 591 | 347 | 10600 |
| h-index (h) | 17 | 17 | 17 | 18 | 18 | 9 | 12 | 12 | 13 | 11 | 11 | 10 | |

**Proceedings of the National Academy of Sciences India Section A - Physical Sciences**

| | | | | | | | | | | | | | |
|---|---|---|---|---|---|---|---|---|---|---|---|---|---|
| No. of papers (P) | 49 | 42 | 40 | 38 | 13 | 68 | 61 | 64 | 83 | 70 | 82 | 156 | 766 |
| No. of uncited papers (U) | 26 | 18 | 25 | 7 | 3 | 19 | 16 | 13 | 15 | 29 | 40 | 97 | 308 |
| Total citation (TC) | 64 | 85 | 39 | 174 | 26 | 352 | 227 | 285 | 296 | 168 | 231 | 181 | 2128 |
| h-index (h) | 4 | 5 | 4 | 7 | 2 | 12 | 8 | 9 | 9 | 6 | 6 | 5 | |